# Stellar Models with Microscopic Diffusion and Rotational Mixing I: Application to the Sun

Brian Chaboyer[1,2], P. Demarque[1,3] and M.H. Pinsonneault[1,4]

## ABSTRACT

The Yale stellar evolution code has been modified to include the combined effects of diffusion and rotational mixing on $^1$H, $^4$He and the trace elements $^3$He, $^6$Li, $^7$Li, and $^9$Be. The interaction between rotational mixing and diffusion is studied by calculating a number of calibrated solar models. The rotational mixing inhibits the diffusion in the outer parts of the models, leading to a decrease in the envelope diffusion by 25 – 50%. Conversely, diffusion leads to gradients in mean molecular weight which can inhibit the rotational mixing. The degree to which gradients in mean molecular weight inhibits the rotational mixing is somewhat uncertain. A comparison to the observed solar oblateness suggests that gradients in the mean molecular weight play a smaller role in inhibiting the rotational mixing previously believed. This is reinforced by the fact that the model with the standard value for the inhibiting effect of mean molecular weight on the rotational mixing depletes no Li on the main sequence. This is in clear contrast to the observations. A reduction in the inhibiting effect of mean molecular weight gradients by a factor of ten leads to noticeable main sequence Li depletion.

*Subject headings:* stars: interiors – stars: rotation – stars: abundances – sun: rotation – sun: interior

[1] Department of Astronomy, and Center for Solar and Space Research, Yale University, Box 208101, New Haven, CT 06520-8101
[2] CITA, 60 St. George St., University of Toronto, Toronto, Ontario, Canada M5S 1A7
  Electronic Mail – I:chaboyer@cita.utoronto.ca
[3] Center for Theoretical Physics, Yale University
[4] present address: Department of Astronomy, Ohio State University, 174 W. 18th Ave., Columbus, OH 43210-1106

## 1. Introduction

In standard stellar models, it is assumed that no material is transported in stellar radiative regions. $^7$Li is a particularly good tracer of particle transport in stellar radiative zones, because it is destroyed at a temperature of $\sim 2.5 \times 10^6$ K. For stars with $M \gtrsim 1.0\ M_\odot$, the Li destruction zone is typically located in the radiative region of the star throughout the pre-main sequence and main sequence evolution. Thus, if no particle transport occurs, then the surface value of Li that we observe should be the initial amount. The solar photospheric Li value is a factor of 200 smaller than that observed in meteorites (Anders & Grevesse 1989), indicating that some particle transport has occurred in the solar radiative regions. A comparison of Li abundances in the Hyades with younger clusters also provides strong observational evidence for main sequence Li depletion (Thorburn et al. 1993; Soderblom et al. 1993).

The most striking evidence of particle transport in stellar radiative regions is found in stars in the temperature range 6400 – 6800 K. These stars are deficient in Li by 1 – 2 orders of magnitude (Boesgaard & Tripicco 1986, Boesgaard & Budge 1988). Other clusters of similar age, and older, also contain a Li gap: Coma, UMa, and NGC 752. However, the younger clusters Pleiades and $\alpha$ Per do *not* show the presence of a Li dip (see discussion by Boesgaard 1991). To date, no stellar evolution models have been able to fully explain the presence of the Li gap, though attempts have been made (Pinsonneault, Kawaler & Demarque 1989; Richer & Michaud 1993). Its existence clearly points to rapid transport in the radiative regions of some stars.

Another difficulty for the standard stellar evolution models was found by recent observations of Pleiades and Hyades stars which demonstrate that a large dispersion exists in the Li abundance for stars of the same mass (Soderblom et al. 1993; Thorburn et al. 1993). In standard stellar models, the amount of $^7$Li depletion depends only on the star's mass, composition and age. Thus, another parameter must govern the depletion of Li.

Together, these observations require that some transport mechanism is operating in stellar radiative regions. In this paper, we examine the hypothesis that this particle transport is caused by microscopic diffusion and rotational instabilities and explore the interactions between the two processes. Previous studies have examined particle transport due to diffusion (eg. Michaud 1986; Proffitt & Michaud 1991; Chaboyer et al. 1992; Richer & Michaud 1993), due to rotational instabilities (Endal & Sofia 1978; Pinsonneault et al. 1989, 1990), and due to the combined effect of diffusion and advection due to meridonal circulation (Charbonneau & Michaud 1991; Charbonneau 1992). In this paper, we study the combined effect of diffusion, and rotation induced transport due to: meridonal circulation, dynamical shear instabilities, the Solberg-Hoiland instability, the Goldreich-Schubert-Fricke instability and the secular shear instability. In contrast to Charbonneau & Michaud, who treated the meridonal circulation as a advection process, we treat meridonal circulation as a diffusion process. This is in agreement with the work of Chaboyer & Zahn (1992), who showed that appropriate rescaling of the turbulent diffusion coefficient allows the combined advection/diffusion process induced by meridonal circulation to be treated solely as a diffusion process.

It is important to note that diffusion will not just affect the surface abundance of $^7$Li, for diffusion will also cause $^4$He to sink relative to hydrogen. Helium diffusion is particularly important in affecting the structure (Vauclair, Vauclair & Pamjatnikh 1974; Bahcall & Pinsonneault 1992; Richer, Michaud & Proffitt 1992) and evolution (Noerdlinger & Arigo 1980; Chaboyer et al. 1992) of a star because helium is very abundant and its diffusion time scale is relatively short.

The rotational models at Yale have been able to reproduce many of the observations of $^7$Li depletion patterns and rotational velocities observed in young cluster stars (Pinsonneault et al. 1990). However, these models ignored microscopic diffusion, which also can have important effects on the calculated particle transport. In particular, the diffusion of $^4$He will lead to gradients in the mean molecular weight. A gradient in mean molecular weight tends to inhibit rotational mixing, as extra energy is required to transport material to regions of different mean molecular weight (Mestel 1953). Thus, diffusion will tend to inhibit the mixing induced by rotation. Conversely, the rotational mixing can inhibit the separation process caused by microscopic diffusion or radiative levitation. When discussing the competing effects of diffusion and rotational mixing, one must keep in mind the key difference in these effects: diffusion is a separation process, with elements sinking with respect to hydrogen; while rotational mixing does not separate elements, it merely mixes them into different regions.

The Sun provides a unique calibration and test of stellar models since its properties (age, chemical composition, luminosity, etc) are known far more accurately than any other star. In addition, helioseis-mology and the solar oblateness provide information regarding the internal rotation of the Sun which is not available for other stars. Finally, by comparing photospheric and meteoritic abundances it is possible to determine the amount of destruction of $^7$Li. The Sun is the only star for which it is possible to make this measurement. For this reason, we start our exploration of the effects of microscopic diffusion and rotational mixing on stellar evolution models by constructing a series of calibrated solar models. It must be stressed however, that in performing this research, we have also considered the constraints from observations of Li abundances and surface rotation velocities in open cluster stars. Observations of open cluster stars of various ages and masses provide provide crucial information regarding the time-scale and mass dependance of $^7$Li depletion which are not available solely from a study of the Sun.

For comparison purposes a number of solar models have been constructed: standard solar models; models which only includes diffusion; a model which only includes rotational mixing; and models which include both rotational mixing and diffusion. This paper is the first in a series which will examine the role of particle transport in stellar radiative regions. Here, we introduce the method which is used to model the particle transport, and concentrate on the interaction between diffusion and rotational mixing by studying the calibrated solar models. Paper II (Chaboyer, Demarque & Pinsonneault 1994) explores the various parameters associated with the stellar models and compares the models to observations of Li and rotation velocities in young cluster stars. A related paper (Chaboyer & Demarque 1994) examines the constraints that observed halo star Li abundances put on particle transport in stellar radiative regions.

In §2. we provide a description of the construction of our solar models, including the treatment of particle in the radiative regions of our models, and list the observed solar parameters which are used to calibrate our models. The calibrated solar models are presented in §3. and the interaction of rotational mixing and microscopic diffusion is discussed. The $^7$Li depletion and surface rotation history of the solar models are shown in §4. Finally, the main results of this paper are summarized in §5.

## 2. Model Construction

### 2.1. Overall Approach

The solar models are constructed by evolving a fully convective, pre-main sequence 1 $M_\odot$ model (with a nominal age of 1 Myr) to the age of the Sun. The initial $X$, $Z$ and $\alpha$ are varied until the solar age model has the solar values of luminosity, radius and $Z/X$. The one dimensional rotating stellar structure equations (see Pinsonneault et al. 1989) are solved using the Henyey relaxation method and are converged to better than 0.1% in each time step. Typically, $\sim 330$ evolutionary time steps (of which 70 were on the main sequence) were required to evolve to a solar model.

After each evolutionary time step, the stability of the angular velocity profile from the previous time step is checked and the diffusion coefficients are calculated (see §2.3.). If the diffusion coefficients are large, then a series of small time steps are taken, with the thermal structure determined by linearly interpolating between the previous and current model. In each time step, the coupled nonlinear diffusion equations for angular momentum and chemical composition (equations (1) and (3)) are solved using an iterative technique, with the convergence criterion set to 0.1%.

All of the solar models presented here match the solar radius and luminosity to within 0.1%, while the surface $Z/X$ was matched to the observed value to within 1.0%. Models which include rotation were required to match the observed solar rotation rate and $^7$Li depletion to within 1.0%. This was achieved by varying the values of $f_c$ and $f_k$ (see equations (3) – (5)).

Our models include the latest physics available to us. The energy producing nuclear reaction rates are from Bahcall & Pinsonneault (1992); reaction rates for the $^6$Li, $^7$Li and $^9$Be are from Caughlan & Fowler (1988); the high temperature opacities from Iglesias & Rogers (1991); the low temperature opacities (below $10^4$ K) are from Kurucz (1991, and private communication); while the surface boundary conditions are determined using Kurucz model atmospheres (Kurucz 1992, and private communication; Howard, 1993). For temperatures above $10^6$ K, a relativistic degenerate, fully ionized equation of state is use. Below $10^6$ K, the single ionization of $^1$H , the first ionization of the metals and both ionizations of $^4$He are taken into account via the Saha equation.

### 2.2. Calibrated Solar Models

A calibrated solar model is a 1 $M_\odot$ stellar model which, at the age of the Sun, has the observed solar radius, luminosity and surface chemical composition. The age of the Sun may be determined by radioactive dating of the oldest meteorites (Tilton 1988), modern theories of star formation, observations of T-Tauri star and pre-main sequence calculations. The best estimate for the age of the Sun is $4.52 \pm 0.4$ Gyr (Guenther *et al.* 1992, Guenther 1989). We have adopted 4.55 Gyr for the solar age in all of our solar models. For the radius, mass and luminosity, we adopt the same values as Guenther *et al.* (1992): $M_\odot = (1.9891 \pm 0.0004) \times 10^{33}$ g; $R_\odot = 6.9598 \times 10^{10}$ cm at $\tau = 2/3$; and $L_\odot = 3.8515 \times 10^{33}$ ergs s$^{-1}$.

The chemical composition of the Sun may be inferred by examining the photospheric abundances in the present Sun and the abundances found in the oldest meteoritics (Anders & Grevesse 1989). For most elements, there is excellent agreement between the two techniques. The exceptions are $^7$Li, $^9$Be and Fe. It now appears that the Fe abundances in the photosphere were incorrect; new measurements give similar abundances to those found in meteorites (Biémont *et al.* 1991; Holweger *et al.* 1991). In our solar models, we adopt the Anders & Grevesse (1989) photospheric heavy element mixture with the meteoritic Fe abundance.

Unfortunately, it is impossible to determine the abundance of $^1$H or $^4$He, and so the mass fraction of $^1$H ($X$) is treated as a free parameter. The mass fraction of $^4$He ($Y$) is determined by requiring the total mass fraction ($X + Y + Z$, where $Z$ is the mass fraction of the heavy element mixture) be equal to one. The Anders & Grevesse photospheric mixture with meteoritic Fe has $Z/X = 0.0267 \pm 0.001$.

There is one other free parameter in a standard solar model, the mixing length, $\alpha$ which is used to describe the convective transport of energy. The outermost layers of the Sun are convective, thus the efficiency of the convective energy transport largely determines the radius of our solar models. The value of $\alpha$ is determined by the requirement that the models have the observed solar radius at the solar age.

In addition to the above surface properties of the Sun, it has recently become possible to determine the location of the base of the convection zone using $p$-mode oscillation frequencies. The depth of the convection zone is found to be $0.713 \pm 0.003 R_\odot$ (Christensen-Dalsgaard, Gough & Thompson 1991). For a given set of input physics (opacities, atmospheres, diffusion, etc), the depth of the convection zone is a fixed quantity and can be considered a prediction of that particular solar model.

For the rotating models, the initial angular momentum distribution must be prescribed. Since the models are evolved starting with a fully convective pre-main sequence model and we assume solid body rotation within convection zones[1], the initial angular momentum distribution is fully described by the initial surface rotation velocity. Studies of T-Tauri indicates that 1 $M_\odot$ stars have rotation periods in the range 1.2 – 12 days and rotation velocities between 8 – 70 km/s (Bouvier 1991, Bouvier *et al.* 1993). The majority of T-Tauri stars have rotation velocities around 10 km/s. It appears that the Sun is over-depleted in $^7$Li compared to other stars of its age. Since the amount of mixing in a star increases as its initial rotational velocity increases, it is likely that the Sun was initially a rapid rotator. For this reason, we have chosen an initial rotation velocity of 30 or 50 km/s for the solar models.

Two additional constraints are imposed on the rotating models. We require that the models have the solar equatorial velocity rate $2.02 \pm 0.04$ km/s (Libbrecht & Morrow 1991) and the observed depletion of $^7$Li. For convenience, we have calibrated the surface rotation velocity of our models to within 1% of 2.0 km/s.

The depletion of $^7$Li is inferred by comparing the cosmic abundance to the photospheric abundance and is a factor of 200 (eg. Duncan 1981). A comparison of the meteoric abundance to the photospheric abundance shows the depletion to be $140^{+40}_{-30}$ (Anders & Grevesse 1989). For most of this work, we have assumed a solar $^7$Li depletion of 200. It is important to realize that in standard solar model, the $^7$Li depletion is very small (factor of 2 – 5). Thus, we do not attempt to fit the $^7$Li depletion with our non-rotating models.

The Sun is unique among stars in that it is possible to obtain information regarding the interior rotation of the Sun. The oblateness of the Sun ($\epsilon$) is a function of the surface solar rotation rate and of the quadrupole moment of the Sun. Since the quadrupole moment is a function of the interior rotation velocity, the oblateness of the Sun constrains the allowed rotation profiles of a solar model. Observations of the oblateness of the Sun are rather difficult. Dicke & Goldenberg (1967) found $\epsilon = (5.0 \pm 0.7) \times 10^{-5}$, a value which implies that general relativity incorrectly predicts the perihelion advance of Mercury. However, Hill & Stebbins (1975) determined $\epsilon = (9.6 \pm 6.5) \times 10^{-6}$, and suggested that the Dicke & Goldenberg observations were hindered by brightness variations between the pole and equator of the Sun. Dicke, Kuhn & Libbrecht (1985) found $\epsilon = (2.0 \pm 0.4) \times 10^{-5}$. Most recently, Sofia, Heaps & Twigg (1994) measured the oblateness using a balloon borne experiment and found $\epsilon = (8.6 \pm 0.9) \times 10^{-6}$.

---

[1] The validity of this assumption is addressed at the end of this sub-section.

This is by far the most accurate measurement of the solar oblateness to date. We have chosen to ignore the Dicke & Goldenberg (1967) results as they are incompatible with the more recent determinations. In using the solar oblateness measurements to test the models, it must be remembered that the oblateness yields information about the gross properties of the rotation profile; different rotation laws may have similar values for the solar oblateness.

In principle, the splitting of the solar $p$-modes contain detailed information on the latitude and depth dependance of the solar angular velocity. Libbrecht (1989) obtained observations of the $\ell = 10 - 60$ $p$-modes, which yield information about the rotation velocity down to $\sim 0.4$ $R_\odot$ (Demarque & Guenther 1988). These observations show that in the equatorial plane, the convection zone essentially rotates as a solid body. As is well known, there is a significant dependence of the surface solar rotation rate on latitude. The latitude dependence of the angular velocity decreases below the surface such that at the bottom the convection zone the angular velocity is independent of latitude (Libbrecht & Morrow 1991) to within $\sim 10\%$. Measurements of low $\ell$ $p$-modes from space (Toutain & Fröhlich 1992) yield information regarding the rotation rate of the deep interior of the Sun and suggest that at $r \sim 0.2$ $R_\odot$, the Sun rotates $\sim 5$ times faster than the surface (Toutain & Fröhlich 1992; Goode, Fröhlich & Toutain 1992). The IRIS network observed the $\ell = 1$ mode (Loudagh et al. 1993) and found a smaller rotational splitting leading to central rotation rates of $\sim 3$ times the surface value. We must caution however, that the weight of the core ($r \leq 0.3$ $R_\odot$) on the total splitting is less than 10%, even these low $\ell$ models (Loudagh et al. 1993). In addition, these results are determined via an inversion procedure, which is sensitive to observational errors and the type of inversion procedure employed (see Goode et al. 1991).

## 2.3. Treatment of Mixing in Radiative Regions

In the radiative regions of a star the transport of angular momentum and chemical composition are treated in a similar manner to Pinsonneault et al. (1989) using the coupled nonlinear equations:

$$\rho r^2 \frac{I}{M} \frac{d\omega}{dt} = \frac{d}{dr}\left[\rho r^2 \frac{I}{M} D_{rot} \frac{d\omega}{dr}\right] \quad (1)$$

$$\rho r^2 \frac{dX_i}{dt} = \frac{d}{dr}\left[\rho r^2 f_m D_{m,1} X_i \right. \quad (2)$$
$$\left. + \rho r^2 (f_m D_{m,2} + f_c D_{rot}) \frac{dX_i}{dr}\right]$$

where $\omega$ is the angular velocity, $X_i$ is the mass fraction of chemical species $i$, $I/M$ is the moment of inertia per unit mass, $D_{m,1}$ and $D_{m,2}$ are derived from the microscopic diffusion coefficients and multiplied by the adjustable parameter $f_m$, $D_{rot}$ is the diffusion coefficient due to rotational instabilities and $f_c$ is an adjustable parameter, which represents the fact that the transport of angular momentum and a chemical species occur on different time-scales. The value of $f_c$ is approximately 0.033 (Chaboyer & Zahn 1992). The value of $f_c$ is varied (within reasonable limits) until the correct $^7$Li depletion for the Sun is obtained. The transport of angular momentum due to the microscopic diffusion is not included, as it is a second order effect.

In all of the models, a small over-mixing zone (0.02 to 0.05 pressure scale heights) below the base of the surface convection zone has been included. This over-mixing zone has been included as many of the estimates for the rotational mixing coefficients go to infinity at the base of the convection zone. The size of the over-mixing zone at the base of the convection zone can be important in determining the amount of pre-main sequence $^7$Li depletion. However, the relatively small over-mixing zones used in these models do not appreciable effect the models.

### 2.3.1. Microscopic Diffusion Coefficients

The microscopic diffusion coefficients of Michaud & Proffitt (1993) are used throughout this paper. For $^1$H and $^4$He, these coefficients are $\sim 20\%$ larger than the Bahcall & Loeb (1990) coefficients. Recent work by Thoul, Bahcall & Loeb (1994) who solve the full system of Burgers equations find agreement with Michaud & Proffitt to within $\lesssim 15\%$. We have introduced an adjustable parameter $f_m$ which multiplies all of the microscopic diffusion coefficients. For the work presented here, we have set $f_m = 0.8$ or 1.0.

This work does not include radiative levitation. This is a good approximation when the atoms are fully ionized. The light elements $^3$He, $^7$Li, and $^9$Be are fully ionized throughout most of the radiative region of our models and so we use the above equations for the microscopic diffusion coefficients for the light elements. It is not clear how important radiative levitation (or diffusion) is for heavier elements. Observations of heavy elements in Hyades stars show similar abundances in stars of different spectral types (Boesgaard & Budge 1988). Thus, we have chosen not to include the effects of microscopic diffusion on any element heavier than $^9$Be .

*2.3.2. Rotational Diffusion Coefficients*

Rotational mixing is triggered by various hydrodynamic instabilities that occur in a rotating star and lead to the bulk transport of matter in the radiative regions of a star (see Zahn 1993 for a recent review). The evolution code at Yale is able to follow the evolution of rotating stars and includes the effects of angular momentum loss due to stellar winds, and the transport of angular momentum and chemical species due to various hydrodynamic instabilities (Pinsonneault *et al.* 1989). It is assumed that convection enforces solid body rotation in the convective regions of a star, so that the rotational instabilities are only effective in the radiative regions.

Rotation induced mixing occurs on two different time-scales: dynamical and secular. The dynamical instabilities have associated time-scales for mixing which are orders of magnitude smaller than other time scales involved in the hydrostatic evolution of a star. As such, if a region is dynamically unstable, the angular momentum distribution is instantaneously readjusted to a point of marginal stability. If a composition gradient existed in the unstable region, then it is reduced by the same amount as the angular velocity gradient. Angular momentum and particle numbers are conserved within the regions subject to the instabilities. In its present form, our stellar evolution code treats two types of rotation-related dynamical instabilities: the dynamical shear instabilities Zahn (1974) and the Solberg-Hoiland instability (cf. Wasiutynski 1946). The stability criterion used are the same as Pinsonneault *et al.* (1989).

Rotation induced instabilities whose time-scale for mixing is comparable to evolutionary time-scales (secular), are treated by solving the coupled diffusion equations (1) and (3). The turbulent diffusion coefficient $D_{rot}$ in these equations is taken to be the product of a velocity estimate $v$ and path length $r$ (Zahn 1993). Note that Pinsonneault *et al.* (1989, 1990) determined the turbulent diffusion coefficient as the product of a velocity estimate and a *local scale height*. Thus, the estimate for the rotational mixing in these models is somewhat different from Pinsonneault *et al.* (1989, 1990). The secular instabilities treated in this work are: meridional circulation (von Zeipel 1924; Eddington 1925; Kippenhahn & Möllenhof 1974), the Goldreich-Schubert-Fricke (GSF) instability (Goldreich & Schubert 1967; Fricke 1968) and the secular shear instability (Zahn 1974). In all cases, the criterion for stability and estimate of the diffusion velocity are calculated in the same manner as Pinsonneault *et al.* (1991). Gradients in the mean molecular weight ($\mu$) inhibit mixing due to meridional circulation (Mestel 1953) and the GSF instability. This effect is taken into account by calculating a fictitious $\mu$ velocity (Mestel 1953) which is subtracted from the velocity estimate for the meridional circulation and GSF circulation.

The amount of mixing induced by the GSF instability is particularly uncertain. For this reason, we have introduced an adjustable parameter, $f_{GSF}$ which multiplies the velocity estimate of James & Kahn (1970, 1971). Values of $f_{GSF}$ greater than one, imply very efficient mixing due to the GSF instability.

In addition, the inhibition of rotational mixing due to gradients in the mean molecular weight is not well understood. As the microscopic diffusion builds up gradients in the mean molecular weight, this is a key parameter when studying the interaction between diffusion and rotational mixing. For this reason, we include a parameter $f_\mu$ which multiplies the velocity estimate for the fictitious $\mu$ current and determines the efficiency of rotational mixing in areas which have a mean molecular weight gradient. Hence, the estimate for the particle transport caused by rotational instabilities contains three parameters: $f_c$, $f_{GSF}$ and $f_\mu$. The parameter $f_c$ is determined by requiring that the $^7$Li depletion in our models match the observed depletion. The parameters $f_\mu$ and $f_{GSF}$ are treated as free parameters in this work.

We have not included angular momentum transport by magnetic fields or internal gravity waves (Press 1981). Either could in principle be important, but both suffer from significant uncertainties in quantifying their impact. Torsional oscillations act to remove angular velocity gradients in a relatively short timescale (see for example Spruit 1987) provided that the magnetic field strengths are allowed to grow arbitrarily large. However, magnetic instabilities could prevent the growth of large internal fields (see Mestel and Weiss 1987), or internal magnetic fields could be decoupled from the envelope (see Charbonneau & MacGregor 1993 for a discussion). We note that both stellar and solar observations provide evidence for differential rotation with depth inside stars, which constrains the role of magnetic fields and waves. Further data on internal stellar rotation, and further advances in theory, will be needed to quantify the role of these potentially important players.

*2.3.3. Angular Momentum Loss*

It is clear from observations of young clusters, and from the fact that the Sun is a slow rotator, that stars lose angular momentum as they evolve (Kraft 1970).

It is likely that this loss is caused by a magnetic stellar wind (Schatzman 1962). Kawaler (1988), following Mestel (1984) developed a general parameterization for the loss of angular momentum due to magnetic stellar winds. The rate of angular momentum loss was proportional to $\omega^{1+4N/3}$, where $N$ is a measure of the magnetic field geometry and may vary from 0 to 2. Since Kawaler's work, it has become clear that such a simple dependance on $\omega$ is not valid; in particular it appears that there is a saturation level in the angular momentum loss (Stauffer & Hartmann 1987; Bouvier 1991) in the pre-main sequence. For this reason we have modified Kawaler's relation to be

$$\frac{dJ}{dt} = f_k K_w \left(\frac{R}{R_\odot}\right)^{2-N} \left(\frac{M}{M_\odot}\right)^{-N/3} \quad (3)$$
$$\cdot \left(\frac{\dot{M}}{10^{-14}}\right)^{1-2N/3} \omega^{1+4N/3} \; ; \quad \omega < \omega_{\rm crit}$$

$$\frac{dJ}{dt} = f_k K_w \left(\frac{R}{R_\odot}\right)^{2-N} \left(\frac{M}{M_\odot}\right)^{-N/3} \quad (4)$$
$$\cdot \left(\frac{\dot{M}}{10^{-14}}\right)^{1-2N/3} \omega \omega_{\rm crit}^{4N/3} \; ; \quad \omega \geq \omega_{\rm crit}$$

where $\omega_{\rm crit}$ introduces a saturation level into the angular momentum loss law, $K_w = 2.036 \times 10^{33} \cdot (1.452 \times 10^9)^N$ in cgs units, $\dot{M}$ is the mass loss rate in units of $10^{-14} \, M_\odot \, {\rm yr}^{-1}$. The dependance on mass loss rate is very weak, for this study we have set it to $2.0 \times 10^{-14} \, M_\odot \, {\rm yr}^{-1}$. Thus, there are 3 adjustable parameters in the wind loss law, $\omega_{\rm crit}$, $N$ and $f_k$. The parameter $N$ is between 0 and 2. The value of $N$ determines the rate of angular momentum loss with time. The value of $f_k$ determines the total amount of the angular momentum loss and it is adjusted until our models have the solar rotation velocity at the solar age. There are no reliable values for $\omega_{\rm crit}$; it is a free parameter in our models.

To be consistent, a global outward drift term should be added to equation (3) in order to take into account the effect of mass loss on element diffusion. However, we have run a few test cases which indicate that, at the level of mass loss considered here ($2.0 \times 10^{-14} \, M_\odot \, {\rm yr}^{-1}$), the mass loss term has little effect on the element diffusion for the masses below $1.3 \, M_\odot$. For this reason, the mass loss term in the element diffusion equation (3) has been ignored.

## 3. Solar Models

### 3.1. Basic Parameters

In order to study the interaction of rotation and diffusion, it is important to separate out the various effects. For this reason, we have constructed a number of calibrated solar models: standard solar models (no rotation or diffusion), a purely rotational model; models which only include diffusion; and a number models which include rotation and diffusion. Although there are a number of adjustable parameters associated with the rotating models, Extensive testing indicates that only $f_\mu$ plays a key role in determining the interaction between rotational mixing (Chaboyer 1993) and diffusion.

The properties of our solar models are shown in Table 1. Each solar model is designated by a two letter code, given in the first column, which will be used to refer to the models throughout the rest of this paper. For each model, the input mixing length, helium abundance ($Y_o$), $Z_o/X_o$ and extent of the overshoot region are shown in columns 2 – 5. Column 6 indicates the constant which multiplies the diffusion coefficients. A value of 0 indicates that diffusion was not included. Columns 7 – 11 give the rotation parameters. A value of 0 in column 7 (the initial rotation velocity) indicates that rotation is not included in that model. Some model results are given in the last two columns. The ratio of the initial $^7$Li abundance to the final surface value is shown in column 12. Finally, column 13 gives the convection zone depth. The different types of models (rotation and diffusion, pure rotation, pure diffusion or standard) are separated from each other by a horizontal line. Discussion of some of the models in Table 1 is deferred to Paper II.

It is clear from Table 1 that the depth of the convection zone matches the observed value ($0.713 \pm 0.003$) if we include rotation and microscopic diffusion. Pure diffusion models have slightly deeper convection zones, and the match to observations is not as good. Models which do not include diffusion (with or without rotation) fail to match the observed convection zone depth, and so might be ruled out as good solar models. We caution, however, that solar models that have no diffusion but have opacities enhanced near the base of the convection zone can match the observed depth of the convection zone (Guzik & Cox 1993).

The present solar rotation curve is a strong function of $f_\mu$. Figure 1 plots the solar rotation curve for the four rotating models which differ in $f_\mu$. All of the calculated solar models show a substantial differen-tial rotation, with the core rotating many times faster than the surface. Below $r = 0.2\ R_\odot$ the rotation rate in model EI (no diffusion, $f_\mu = 0.1$) rises steeply to match that of model AM (microscopic diffusion, $f_\mu = 0.1$). Above $r = 0.2\ R_\odot$, the rotation profile in model EI is nearly identical to model CO ($f_\mu = 0.01$). This clearly shows the inhibiting effects of microscopic diffusion on the rotational induced mixing in the outer layers of the Sun. Models with microscopic diffusion require a value of $f_\mu$ ten times smaller in order to achieve a similar rotation profile in the outer layers to the pure rotation models.

Table 1
Solar Models

| | Input Parameters | | | | | | | | | | Results | |
|---|---|---|---|---|---|---|---|---|---|---|---|---|
| RUN (1) | Mixing Length (2) | $Y_o$ (3) | $Z_o/X_o$ (4) | Overshoot ($H_p$) (5) | $f_m$[a] (6) | $V_i$[b] (km/s) (7) | N[c] (8) | $f_{GSF}$[d] (9) | $f_\mu$[e] (10) | $\omega_{crit}$[f] ($10^{-5}s^{-1}$) (11) | $Li_i/Li_f$ (12) | CZ depth[g] ±0.0009 (13) |
| BF | 1.841 | 0.28167 | 0.02768 | 0.05 | 1.0 | 30 | 1.5 | 10 | 1.00 | 3.0 | 199.2 | 0.7096 |
| AM | 1.832 | 0.28155 | 0.02768 | 0.05 | 1.0 | 30 | 1.5 | 10 | 0.10 | 3.0 | 199.0 | 0.7101 |
| CO | 1.827 | 0.28155 | 0.02768 | 0.05 | 1.0 | 30 | 1.5 | 10 | 0.01 | 3.0 | 200.9 | 0.7109 |
| IF | 1.827 | 0.28180 | 0.02769 | 0.02 | 1.0 | 30 | 1.5 | 10 | 0.01 | 3.0 | 199.6 | 0.7110 |
| HC | 1.827 | 0.28175 | 0.02769 | 0.05 | 1.0 | 50 | 1.5 | 10 | 0.01 | 3.0 | 200.6 | 0.7100 |
| UU | 1.838 | 0.28155 | 0.02768 | 0.05 | 1.0 | 30 | 2.0 | 10 | 0.10 | 3.0 | 200.0 | 0.7090 |
| DQ | 1.837 | 0.28145 | 0.02767 | 0.05 | 1.0 | 50 | 2.0 | 10 | 0.10 | 3.0 | 199.0 | 0.7098 |
| RK | 1.828 | 0.28185 | 0.02769 | 0.05 | 1.0 | 30 | 2.0 | 1.0 | 0.10 | 3.0 | 202.1 | 0.7097 |
| JF | 1.804 | 0.28171 | 0.02755 | 0.02 | 0.8 | 30 | 1.5 | 10 | 0.01 | 3.0 | 199.4 | 0.7141 |
| QG | 1.800 | 0.28199 | 0.02756 | 0.02 | 0.8 | 30 | 1.5 | 1.0 | 0.01 | 3.0 | 199.4 | 0.7137 |
| SF | 1.806 | 0.28174 | 0.02755 | 0.02 | 0.8 | 30 | 1.5 | 10 | 0.01 | 1.5 | 199.8 | 0.7125 |
| TJ | 1.806 | 0.28174 | 0.02755 | 0.02 | 0.8 | 30 | 1.8 | 10 | 0.01 | 1.5 | 199.2 | 0.7129 |
| VN | 1.806 | 0.28207 | 0.02757 | 0.02 | 0.8 | 30 | 1.5 | 1.0 | 0.01 | 1.5 | 200.2 | 0.7126 |
| WE | 1.806 | 0.28187 | 0.02756 | 0.02 | 0.8 | 30 | 1.5 | 1.0 | 0.01 | 1.5 | 99.35 | 0.7130 |
| EI | 1.725 | 0.28125 | 0.02686 | 0.05 | 0.0 | 30 | 1.5 | 10 | 0.10 | 3.0 | 200.1 | 0.7241 |
| FC | 1.856 | 0.28176 | 0.02769 | 0.05 | 1.0 | 0 | — | — | — | — | 4.747 | 0.7089 |
| KB | 1.862 | 0.28176 | 0.02769 | 0.02 | 1.0 | 0 | — | — | — | — | 3.648 | 0.7078 |
| XB | 1.856 | 0.28176 | 0.02769 | 0.25 | 1.0 | 0 | — | — | — | — | 202.4 | 0.7094 |
| XC | 1.856 | 0.28176 | 0.02769 | 0.24 | 1.0 | 0 | — | — | — | — | 148.5 | 0.7094 |
| ND | 1.835 | 0.28179 | 0.02756 | 0.02 | 0.8 | 0 | — | — | — | — | 3.358 | 0.7117 |
| GD | 1.728 | 0.28149 | 0.02687 | 0.05 | 0.0 | 0 | — | — | — | — | 2.999 | 0.7246 |
| LA | 1.728 | 0.28149 | 0.02687 | 0.02 | 0.0 | 0 | — | — | — | — | 2.461 | 0.7246 |
| PA | 1.728 | 0.28149 | 0.02687 | 0.00 | 0.0 | 0 | — | — | — | — | 2.151 | 0.7246 |

[a] Constant which multiplies the microscopic diffusion coefficients of Michaud & Proffitt 1993.

[b] Initial rotation velocity given to the models on the pre-main sequence.

[c] Power law index for the wind loss law.

[d] Constant which multiplies the estimate for the GSF diffusion coefficient.

[e] Constant which determines the efficiency of rotational mixing in the presence of mean molecular weight gradients. Small values of $f_\mu$ imply efficient mixing.

[f] Saturation level in the angular momentum loss law.

[g] Depth of the surface convection zone in units of $R_\odot$. In the Sun, the surface convection zone has a depth of $0.713 \pm 0.003\ R_\odot$.

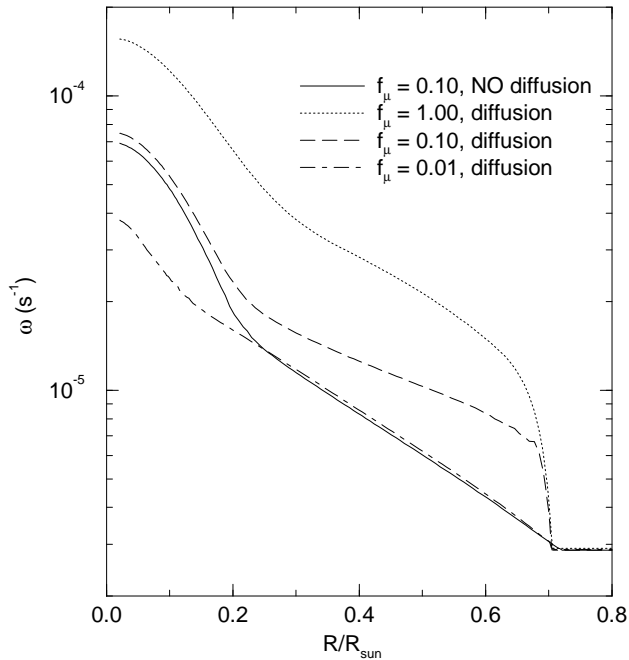

Fig. 1.— Present day solar rotation curve for the rotating models EI (no diffusion), BF ($f_\mu = 1.0$), AM ($f_\mu = 0.1$) and CO ($f_\mu = 0.01$).

Models with $f_\mu = 0.01$ have the lowest core rotation rates, typically a factor of $\sim 12$ faster than the surface. Models with $f_\mu = 0.1$ have core rotation rates $\sim 25$ times faster than the surface. Such high core rotation rates may be ruled out by the observed splittings of the lowest $\ell$ p-modes (Toutain & Fröhlich 1992; Loudagh et al. 1993) which suggest that the core of the Sun is rotating $\sim 4$ times faster than the surface. The lower core rotation rate inferred by Toutain & Fröhlich and Loudagh may be a resolution problem, for even the lowest $\ell$ modes are weighted such that they are ineffective at probing the rotation rate at a radius of $\lesssim 0.2\ R_\odot$. At $r = 0.2\ R_\odot$, models with $f_\mu = 0.01$, $0.1$ and $1.0$ have rotation rates which are $\sim 5$, $8$ and $23$ times faster than the surface rate respectively. Thus, the $f_\mu = 1.0$ case appears to be ruled out on the basis of its extremely high core rotation rate. We note however, that if an angular momentum transport process which does not transport particles (such as magnetic fields, or waves) is active in the Sun, then it is not possible to rule out the $f_\mu = 1.0$ case. The $f_\mu = 0.01$ and $f_\mu = 0.1$ are in reasonable agreement with the observations of the core rotation of the Sun.

Observations of the $\ell = 10 - 60$ yield information of the rotation velocity between $r = 0.4 - 0.9\ R_\odot$. Inversions of these observations indicate that the Sun rotates as a solid body over this range in radii (Libbrecht & Morrow 1991, Goode et al. 1991). All of our models contain a fairly substantial amount of differential rotation between the base of the convection zone and $r = 0.4\ R_\odot$. At $r = 0.4\ R_\odot$, model CO ($f_\mu = 0.01$) has a rotation rate 2.9 times faster than the surface rate. Thus, none of our models agree with the rotation profile inferred by inversions of the $\ell = 10 - 60$ p-modes. However, the inversion process is sensitive to observational errors, and the type of inversion procedure employed (see Goode et al. 1991). It is clear from observations of surface rotation rates in subgiants (Noyes et al. 1984; Pinsonneault et al. 1989; Demarque & Guenther 1988) and in horizontal branch stars (Peterson 1985; Pinsonneault, Deliyannis & Demarque 1991) that a reservoir of angular momentum must be buried inside a star, for these stars rotate substantially faster than what can be accounted for by assuming a solid body rotation on the main sequence. For this reason we do not rule out any of our models based on their rotation profiles. However, it appears that a more efficient angular momentum mixing mechanism is present in the Sun, which is not present in our models. Magnetic fields have been ignored in this study. Large magnetic fields in the radiative region of the Sun have the potential to efficiently mix angular momentum, and lead to a flat solar rotation curve (see Charbonneau & MacGregor 1993).

The quadrupole moments and the associated solar oblateness have been calculated using the method of Fricke (1968, see Sweet 1950 for the original derivation) modified to take into account the solar differential rotation in latitude (Dicke 1970). Model BF ($f_\mu = 1$) has $\epsilon = 2.8 \times 10^{-5}$, which is $3\sigma$ above the observation of Hill & Stebbins (1975), $2\sigma$ above the observations of Dicke et al. (1985) and $> 27\sigma$ above the observations of Sofia et al. (1994). Thus, model BF is ruled out as a viable solar model on the basis of its high quadrupole moment. Model AM ($f_\mu = 0.1$) has $\epsilon = 9.6 \times 10^{-6}$ which is in agreement with the measurements of Hill & Stebbins (1975) and Sofia et

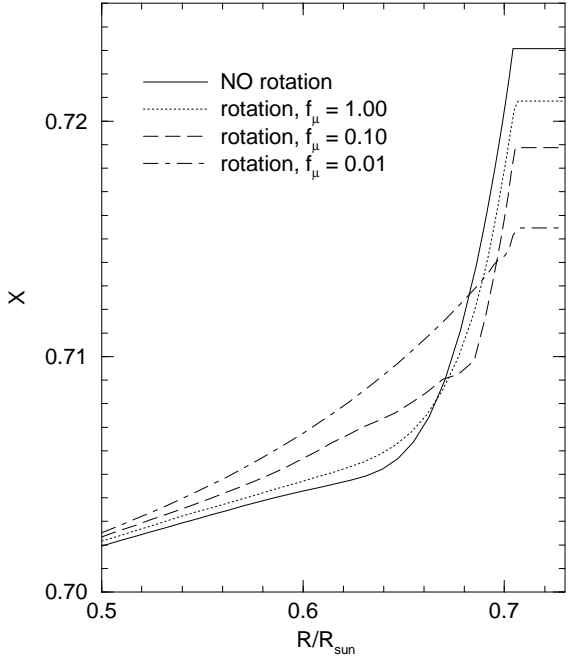

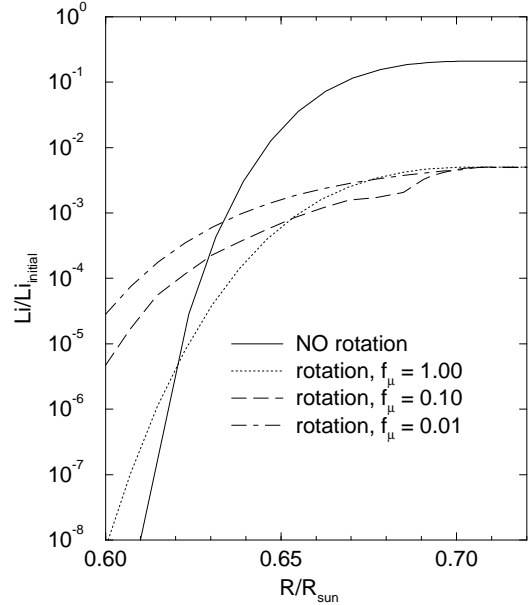

Fig. 2.— The radial distribution of the $^1$H mass fraction ($X$) near the base of the convection zone in solar models which include microscopic diffusion (FC – no rotation; BF – $f_\mu = 1.0$; AM – $f_\mu = 0.1$; and CO – $f_\mu = 0.01$).

Fig. 3.— The radial distribution of the $^7$Li divided by its initial value near the base of the convection zone in solar models which include microscopic diffusion (FC – no rotation; BF – $f_\mu = 1.0$; AM – $f_\mu = 0.1$; and CO – $f_\mu = 0.01$).

*al.* (1994). A similar statement holds for model CO ($f_\mu = 0.01$).

It is clear from Figure 1 that models which include diffusion and a large value of $f_\mu$ show a distinct discontinuity in $\omega$ at the base of the convection zone. This is due to the large gradient in $\mu$ that develops due to the diffusion of $^4$He out of the convection zone. This $\mu$ gradient effectively halts the rotational mixing in models with $f_\mu = 1.0$ and $0.1$. This may be seen in Figure 2 where we have plotted the mass fraction of $^1$H ($X$) near the base of the convection zone for four models which include microscopic diffusion. In contrast to the high $f_\mu$ cases, model CO ($f_\mu = 0.01$) has a relatively small gradient in $X$ at the base of the convection zone ($r = 0.71\,R_\odot$). This is due to the fact that the rotational mixing is able to smooth out some of the gradients composition caused by diffusion. Note that by $r = 0.5\,R_\odot$, the hydrogen profile is very similar among the models. Figure 3 plots the radial profile of the $^7$Li abundance (divided by its initial abundance) for the same models. The rotating models have been calibrated such that the surface $^7$Li abundance is 1/200th of its original value. The region of $^7$Li destruction is more extended in the rotating models than in the pure diffusion model due to the slow timescale of the rotational mixing.

### 3.2. Effect of Rotation on the Microscopic Diffusion

When rotational mixing is active, it tends to inhibit the sorting effect of diffusion. This effect is best exemplified by examining the change in time of the surface $^4$He abundance in calibrated models (Figure 4). It is best to concentrate on the surface abundance of $^4$He because it is only influenced by particle transport; no nuclear destruction or creation contributes to its change. Figure 4 clearly demonstrates the interaction of rotational mixing with microscopic diffusion, and the effects of $f_\mu$. In model BF (with $f_\mu = 1.0$) no rotational mixing occurs near the surface after 1.5 Gyr and the rate of $^4$He depletion is virtually identical to model FC (which has no rotational mixing). In contrast, in model AM ($f_\mu = 0.1$) the rate of $^4$He depletion does not match model FC until $\sim 3.5$ Gyr. The rate of $^4$He depletion is smallest (and relatively constant, at rate $\sim 75\%$ of the pure diffusion model) in model CO ($f_\mu = 0.01$). This suggests that in models with small $f_\mu$, the rotational mixing leads to an 'effective' microscopic diffusion coefficient (used to derive $D_{m,1}$ and $D_{m,2}$) which is about 30% smaller than the calculated coefficients near the surface of the star. Since the microscopic diffusion in the outer part of our model causes a change in the radius of the models, the inhibiting effect of rotational mixing has important implications for effective temperatures of our stellar models.

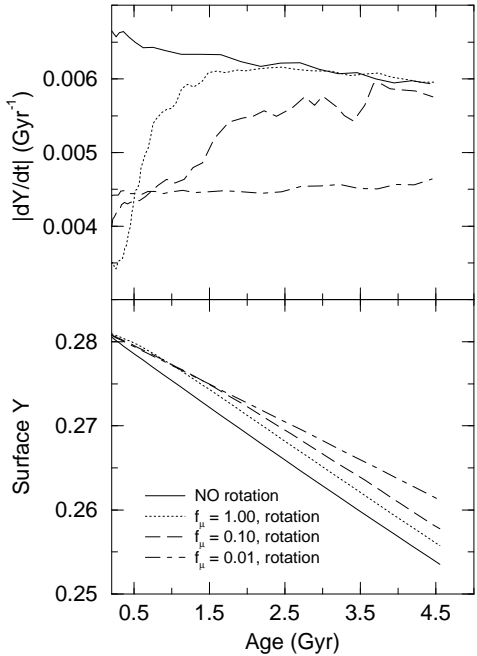
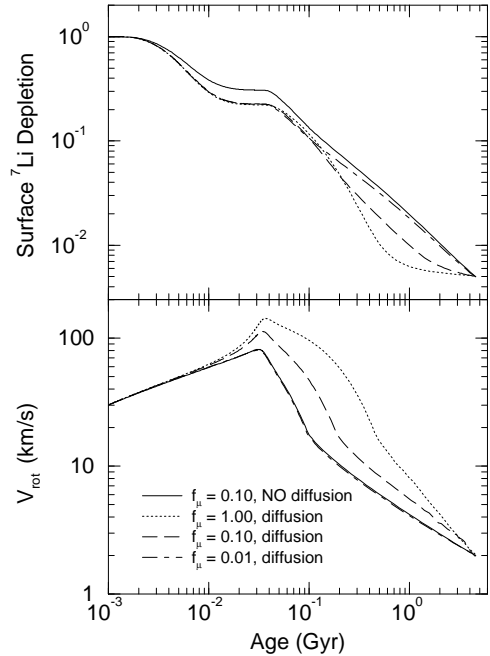

Fig. 4.— The lower panel shows the mass fraction of $^4$He ($Y$) at the surface of the Sun as a function of time for models which include diffusion (FC – no rotation; BF – $f_\mu = 1.0$; AM – $f_\mu = 0.1$; and CO – $f_\mu = 0.01$). The upper panel shows the rate of change in $Y$.

Fig. 5.— The surface rotation velocity (lower panel) and $^7$Li depletion as a function of time for models which include rotational mixing (EI – no diffusion; BF – $f_\mu = 1.0$; AM – $f_\mu = 0.1$; and CO – $f_\mu = 0.01$)

## 4. $^7$Li Depletion and Rotational History of the Solar Models

The details of how the models arrive at their present state varies considerably. It is instructive to examine the variation in time of the surface rotation velocity and $^7$Li as these two quantities are readily observed in other stars. Figure 5 plots the surface rotation velocity ($v$) and $^7$Li depletion as a function of age for four models which include rotation. When examining this plot, it is important to remember that *calibrated* solar models have been calculated; the rotational mixing efficiency ($f_c$) and wind constant ($f_k$) have been adjusted to yield the observed surface rotation rate and $^7$Li abundance of the present day Sun.

On the main sequence, the rate of decline in the surface $v$ is steeper for models with large $f_\mu$. This is because gradients in $\mu$ inhibit the rotational mixing and decouple the core and the envelope. Thus, in models with large $f_\mu$ only the envelope spins down. This means that our calibrated models require a smaller wind constant, implying that the models with large $f_\mu$ spin up more during the pre-main sequence phase of evolution. Hence, beyond $\sim 10$ Myr, models with large $f_\mu$ have larger surface rotation velocities. It is interesting to note that the $f_\mu = 0.01$ model has a nearly identical rotation velocity history to the $f_\mu = 0.10$ pure rotation model.

The effect of $f_\mu$ on the surface $^7$Li abundance is equally striking. In model BF ($f_\mu = 1.0$), the $^7$Li abundance is virtually constant on the main sequence, implying that very little rotational induced mixing occurs. *However, observations of Li in the Pleiades and Hyades show that main sequence Li depletion has occurred* (Soderblom et al. 1993; Thorburn et al. 1993). The $f_\mu = 0.10$ and $0.01$ models show significant main sequence Li depletion. Comparisons to cluster observations are somewhat complex, as allowance must be made for a range in initial rotation velocities. Most T-Tauri stars have rotation velocities near 10 km/s, though some have rotation velocities as high as 70 km/s (Bouvier et al. 1993). To compare to the open cluster observations, Li destruction isochrones were calculated by evolving stellar models with masses of $0.6, 0.7, \ldots 1.3\ M_\odot$ and initial rotation velocities of 10 and 30 km/s. In all other respects, these models used parameters identical to model BF ($f_\mu = 1.0$). The $^7$Li destruction isochrones are compared to $^7$Li observations in the Hyades (age 700 Myr) and M67 (age 4.5 Gyr) in Fig-ure 6. At the age of the Hyades, the 10 km/s Li destruction isochrones lie somewhat below the mean trend of the observations. The 30 km/s isochrones lie well below the mean trend. These models predict a much larger dispersion (at a give $T_{eff}$) than is observed in the Hyades. The lack of main-sequence $^7$Li depletion causes the 10 km/s models to have much more $^7$Li than is observed in older clusters, such as M67. The lack of main-sequence $^7$Li depletion clearly rules out the $f_\mu = 1.0$ case, implying that canonical estimates for the amount of inhibition in rotational mixing are too large. Reducing the inhibiting effect of mean molecular weights by a factor of ten, leads to noticeable main sequence $^7$Li depletion.

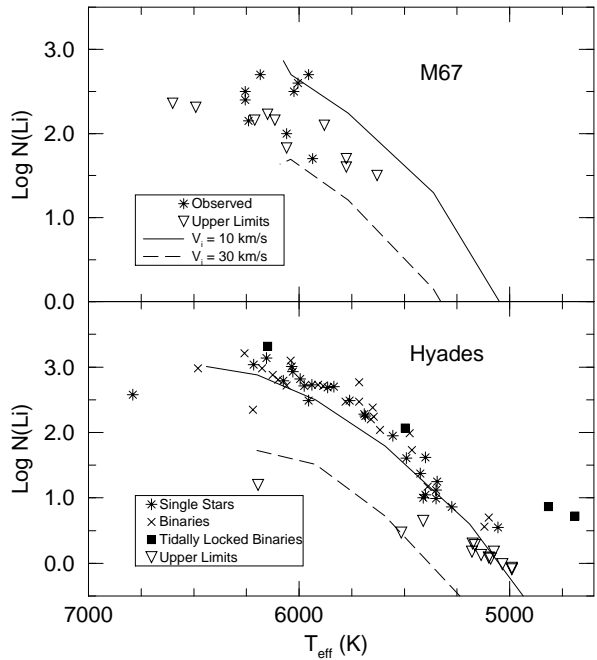

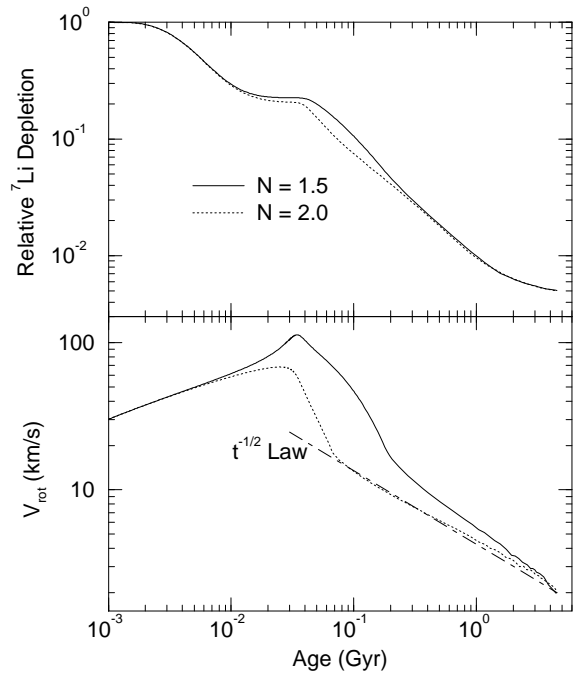

Fig. 6.— Li destruction isochrones for the $f_\mu = 1.0$ BF case are compared to the Hyades (lower panel) and M67 (upper panel) observations. Two sets of Li isochrones are shown – the solid line is for models with an initial rotation velocity of 10 km/s, while the dashed line represents the Li isochrone for a family of models with an initial rotation velocity of 30 km/s. It is clear that the 30 km/s isochrones are too low at the age of the Hyades (700 Myr), while the 10 km/s isochrones are too high at the age of M67 (4 Gyr). This is due to the fact that the models have little main sequence Li depletion. The observations are from Thorburn *et al.* (1993) for the Hyades, and from Hobbs & Pilachowski (1986) and Spite *et al.* (1987) for M67.

Fig. 7.— Effect of changing the power law index $N$ in the wind loss law on the surface $V$ and $^7$Li abundance. Models AM ($N = 1.5$) and UU ($N = 2$) are shown. A Skumanich (1972) $t^{-1/2}$ spin down law is shown in the bottom panel.

Figure 7 illustrates the effect of changing the power law index $N$ in the wind loss law on the models. As expected, models with higher $N$ suffer significantly more angular momentum loss. As a consequence, models with high $N$ do not spin up as much, and so suffer less angular momentum loss on the main sequence. Thus, the surface $^7$Li depletion occurs somewhat earlier in models with high $N$. We have schematically illustrated a Skumanich (1972) $t^{-1/2}$ spin down law on the $V_{rot}$ plot. It is clear that model AM ($N = 1.5$, $f_\mu = 0.1$) spins down too rapidly on the main sequence. Thus, it is ruled out as a viable solar model. Model UU has the maximum $N$ allowed (2) and follows the $t^{-1/2}$ law reasonably well. In Figure 8 the sensitivity of the models to changing the saturation level in the angular momentum loss law is demonstrated. Models with low $\omega_{crit}$ spin up more doing the pre-main sequence and retain substantially faster rotation rates until an age of $\sim 700$ Myr. This is a natural consequence of lowering $\omega_{crit}$, above which the power law dependence of the loss on $\omega$ is lowered. We note that past $\sim 1$ Gyr, the surface rotation velocities of these models are nearly identical. The $^7$Li depletion is somewhat different between the ages of $\sim 60 - 200$ Myr, which suggests that the observations in the Pleiades (age $\sim 70$ Myr) may be able to distinguish between these models.

In Figure 9 we show the consequence of varying the initial rotation velocity on the $^7$Li depletion and surface rotation velocity. For these calibrated models, a higher initial velocity implies a larger wind constant as we require that the models rotate at 2 km/s at the solar age. During the early pre-main sequence phase ($t \lesssim 25$ Myr) the difference in surface rotation velocity remains constant. However, once the star starts to spin down, the model with a higher initial rotation velocity loses angular momentum at a faster rate due to the higher wind constant. The $^7$Li depletion occurs somewhat earlier in models with higher initial rotation velocities, though not by an measurable amount.

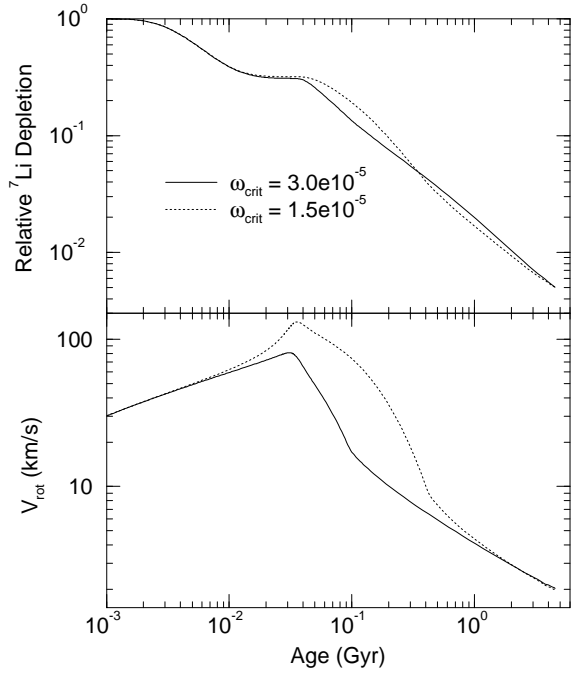

Fig. 8.— Effect of changing the saturation level in the angular momentum loss law on the surface $V$ and $^7$Li abundance. Models JF ($\omega_{\rm crit} = 3 \times 10^{-5}\ s^{-1}$) SF ($\omega_{\rm crit} = 1.5 \times 10^{-5}\ s^{-1}$) are shown.

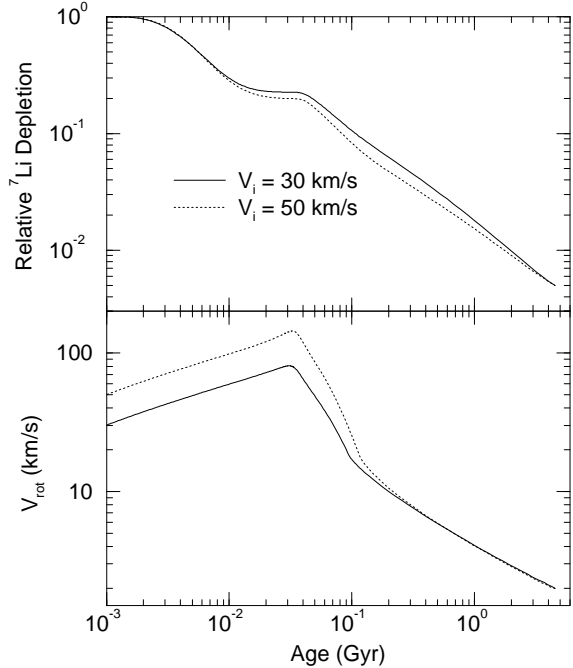

Fig. 9.— Effect of changing the initial rotation velocity on the surface velocity $V$, and $^7$Li abundance. Models CO ($V_i = 30$ km/s) and HC ($V_i = 50$ km/s) are shown.

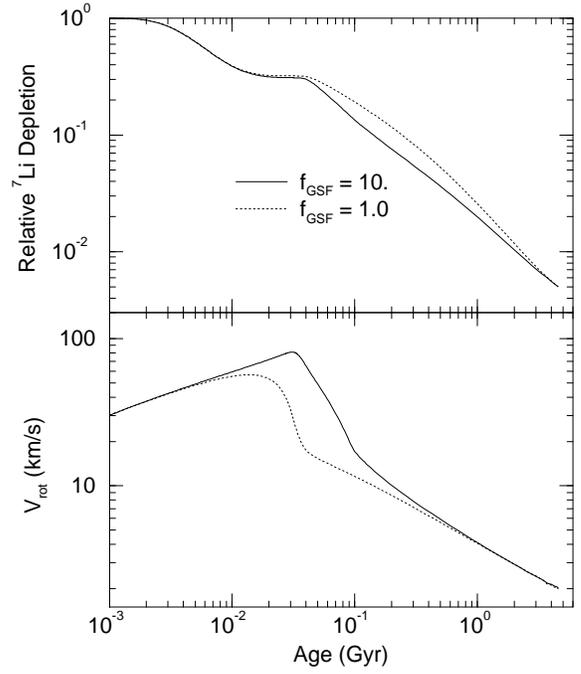

Fig. 10.— Effect of changing the efficiency of the GSF instability in mixing material. Models JF ($f_{\rm GSF} = 10$), and QG ($f_{\rm GSF} = 1.0$) are shown.

Finally, Figure 10 shows the effect of modifying the GSF coefficients. Models with small $f_{\rm GSF}$ do not spin up as much on the pre-main sequence and have lower surface rotation velocities through most of their pre-main sequence and early main sequence evolution ($10 \lesssim t \lesssim 300$ Myr). This is due to the fact that when the GSF instability is inefficient, the main sequence spin down of the star is slow, which requires that the star not spin up too much during the pre-main sequence (recall that we are examining solar models which have been calibrated to yield the surface solar rotation rate at the solar age). The depletion of $^7$Li is delayed when $f_{\rm GSF}$ is lowered, due to the decreased efficiency of the GSF instability in mixing material. It is important to note that when the efficiency of the GSF instability is lowered, the mixing in higher mass stars is lowered relative to lower mass stars.

It is clear from Figures 7 – 10 that our models display a complex behaviour in terms of surface $^7$Li depletion and rotation velocities. However, a few general conclusions may be drawn. The power law index in the wind law ($N$) determines the rate of spin down on the main sequence. By varying $N$, it is possible to achieve a Skumanich $t^{-1/2}$ (1972) spin down on the main sequence. A low value for the angular momentum saturation level leads to substantial higher pre-main sequence and ZAMS rotation velocities. Models with different initial rotation velocities have nearly identical rotation velocities by an age of $\sim 150$ Myr.

## 5. Summary

A number of solar models which include the combined effects of microscopic diffusion and rotationally induced mixing in the radiative regions of the model have been calculated. The models suffer angular momentum loss from the surface. These are the first solar models which included all three of these effects. The models have been calibrated to match the observed solar radius, luminosity, surface $Z/X$, surface rotation velocity, and $^7$Li depletion.

Helioseismology is able to determine the depth of the solar convection zone (Christensen-Dalsgaard et al. 1991) to be $0.713 \pm 0.003~R_\odot$. Solar models which include rotation and microscopic diffusion have convection zone depths of $0.710~R_\odot$, providing a good match to the observed depth. Models which do not include diffusion have convection zones which are too shallow compared to observations. This has been found by other workers, who note that the convection zone depth in standard models may be made to agree with the observations by enhancing opacities near the base of the convection zone (Guzik & Cox 1993).

Models which include microscopic diffusion build up gradients in the mean molecular weight near the base of the convection zone. Such gradients can inhibit the rotationally induced mixing. The sensitivity of rotational mixing to gradients in the mean molecular weight is somewhat uncertain; we have parameterized it by the non-dimensional multiplicative factor $f_\mu$. A large $f_\mu$ implies a strong inhibiting factor due to gradients in mean molecular weight. The model with $f_\mu = 1.0$ (the standard value, given by Kippenhahn 1974) has very high internal rotation velocities which cause a large solar oblateness $\epsilon = 2.8 \times 10^{-5}$. Such a high value of the solar oblateness is ruled out by observations (Sofia et al. 1994). Models with $f_\mu \lesssim 0.1$ are in agreement with the oblateness measurements. A further problem with assuming a high $f_\mu$ is that the model with $f_\mu = 1$ depletes virtually no $^7$Li on the main sequence. This is in clear contrast with the observation of Li in the Pleiades and Hyades (Thorburn et al. 1993; Soderblom et al. 1993). Thus, it is clear that gradients in the mean molecular weight inhibit the rotational mixing by a smaller amount than previously thought.

The rotational mixing inhibits the microscopic diffusion settling in the outer parts of the model. An examination of the surface $^4$He abundance (Figure 4) reveals the amount of inhibition. Models which do not include rotational mixing require their microscopic diffusion coefficients (used to derive $D_{m,1}$ and $D_{m,2}$) to be multiplied by $0.5 - 0.8$ in order to obtain a similar depletion in the surface $^4$He as the models which include rotational mixing. Microscopic diffusion operates relatively unhindered in the core of the model. This may explain the findings of Chaboyer et al. (1992) that the amount of age reduction in globular clusters caused by diffusion depended on the method used. Isochrone fitting and $\Delta(B-V)$ methods suggested age reduction of $2-3$ Gyr, while the $\Delta V(TO-HB)$ method suggested an age reduction of $0.5-1$ Gyr. Both isochrone fitting and $\Delta(B-V)$ depend on the color (hence, radius) of the models, while $\Delta V(TO-HB)$ depends on the luminosity of the models. The amount of microscopic diffusion in the outer part of the star determines the change in radius (hence, color) in the models. The amount of diffusion in the core determines the effect of microscopic diffusion on the luminosity. Thus, the inhibiting effect of rotational mixing in the outer layers of the star is likely to reduce the age reduction found by isochrone fitting and $\Delta(B-V)$ by a factor of two while the age reduction found by $\Delta V(TO-HB)$ is unchanged.

Observations of splittings in $\ell = 10-60$ $p$-modes suggest that between the base of the convection zone and $r = 0.4~R_\odot$ the Sun rotates as a solid body (Libbrecht & Morrow 1991). Our model with $f_\mu = 0.01$ has the least amount of differential rotation of our models. At $r = 0.4~R_\odot$ it has an angular velocity 2.9 times higher than the surface, which contradicts the solid body rotation inferred by inversions of the $\ell = 10-60$ $p$-modes. This suggests that the transport of angular momentum is more efficient in the Sun than what we have modeled. We note however, that it is clear from the relatively rapid rotation of subgiant stars (Pinsonneault et al. 1989) and of horizontal branch stars (Pinsonneault et al. 1991) that stars must have large reservoir of angular momentum which is transported to the surface once the stars leave the main sequence. All of the solar models have a core which rotates substantially faster than the surface.

The observations of the splitting in low $\ell$ $p$-modes by Toutain & Fröhlich (1992) and Loudagh *et al.* (1993) suggest that the rotation rate is $\sim 4$ times faster than the surface at $r \approx 0.2\ R_\odot$. At a similar depth, the $f_\mu = 0.1$ and 0.01 cases have rotation rates 8 and 5 times the surface rate respectively. Thus, they are in reasonable agreement with the splittings in the low $\ell$ $p$-mode observations.

B.C. would like to thank Sabatino Sofia and Douglas Duncan, who provided him with their thoughtful comments on his thesis, which this paper is drawn from. In addition, B.C. is grateful to Charles Proffitt who provided him with detailed information regarding the microscopic diffusion coefficients used in this work. The comments of the anonymous referee improved the presentation of this paper. Research supported in part by NASA grants NAG5–1486, NAGW–2136, NAGW–2469 and NAGW-2531.